\documentstyle[twoside]{article}

\catcode`\@=11
\long\def\@makefntext#1{
\protect\noindent \hbox to 3.2pt {\hskip-.9pt  
$^{{\eightrm\@thefnmark}}$\hfil}#1\hfill}		

\def\@makefnmark{\hbox to 0pt{$^{\@thefnmark}$\hss}}	
	
\def\ps@myheadings{\let\@mkboth\@gobbletwo
\def\@oddhead{\hbox{}
\rightmark\hfil\eightrm\thepage}   
\def\@oddfoot{}\def\@evenhead{\eightrm\thepage\hfil
\leftmark\hbox{}}\def\@evenfoot{}
\def\sectionmark##1{}\def\subsectionmark##1{}}

\oddsidemargin=\evensidemargin
\newcounter{sectionc}\newcounter{subsectionc}\newcounter{subsubsectionc}
\renewcommand{\section}[1] {\vspace{12pt}\addtocounter{sectionc}{1} 
\setcounter{subsectionc}{0}\setcounter{subsubsectionc}{0}\noindent 
	{\tenbf\thesectionc. #1}\par\vspace{5pt}}
\renewcommand{\subsection}[1] {\vspace{12pt}\addtocounter{subsectionc}{1} 
	\setcounter{subsubsectionc}{0}\noindent 
	{\bf\thesectionc.\thesubsectionc. {\kern1pt \bfit #1}}\par\vspace{5pt}}
\renewcommand{\subsubsection}[1] {\vspace{12pt}\addtocounter{subsubsectionc}{1}
	\noindent{\tenrm\thesectionc.\thesubsectionc.\thesubsubsectionc.
	{\kern1pt \tenit #1}}\par\vspace{5pt}}
\newcommand{\nonumsection}[1] {\vspace{12pt}\noindent{\tenbf #1}
	\par\vspace{5pt}}

\newcommand{\textlineskip}{\baselineskip=13pt}
\newcommand{\smalllineskip}{\baselineskip=10pt}

\def\eightcirc{
\begin{picture}(0,0)
\put(4.4,1.8){\circle{6.5}}
\end{picture}}
\def\eightcopyright{\eightcirc\kern2.7pt\hbox{\eightrm c}} 

\newcommand{\copyrightheading}[1]
	{\vspace*{-2.5cm}\smalllineskip{\flushleft
        {\footnotesize Quantum Aspects of Beam Physics, Monterey, Ca, USA,
        4-9 January 1998,\\
        ed. Pisin Chen, pp. 761-763#1}\\
        {\footnotesize $\eightcopyright$\ 1999 by World Scientific Publishing
         Co. Pte. Ltd.}\\
	 }}

\def\abstracts#1#2#3{{
	\centering{\begin{minipage}{4.5in}\baselineskip=10pt\footnotesize
	\parindent=0pt #1\par 
	\parindent=15pt #2\par
	\parindent=15pt #3
	\end{minipage}}\par}} 

\newcommand{\bibit}{\nineit}

\renewenvironment{thebibliography}[1]
	{\frenchspacing
	 \ninerm\baselineskip=11pt
	 \begin{list}{\arabic{enumi}.}
        {\usecounter{enumi}\setlength{\parsep}{0pt}     
	 \setlength{\leftmargin 12.7pt}{\rightmargin 0pt} 
         \setlength{\itemsep}{0pt} \settowidth
	{\labelwidth}{#1.}\sloppy}}{\end{list}}

\newcounter{itemlistc}
\newcounter{romanlistc}
\newcounter{alphlistc}
\newcounter{arabiclistc}

\def\@citex[#1]#2{\if@filesw\immediate\write\@auxout
	{\string\citation{#2}}\fi
\def\@citea{}\@cite{\@for\@citeb:=#2\do
	{\@citea\def\@citea{,}\@ifundefined
	{b@\@citeb}{{\bf ?}\@warning
	{Citation `\@citeb' on page \thepage \space undefined}}
	{\csname b@\@citeb\endcsname}}}{#1}}

\newif\if@cghi
\def\cite{\@cghitrue\@ifnextchar [{\@tempswatrue
	\@citex}{\@tempswafalse\@citex[]}}
\def\citelow{\@cghifalse\@ifnextchar [{\@tempswatrue
	\@citex}{\@tempswafalse\@citex[]}}
\def\@cite#1#2{{$\null^{#1}$\if@tempswa\typeout
	{IJCGA warning: optional citation argument 
	ignored: `#2'} \fi}}

\def\@refcitex[#1]#2{\if@filesw\immediate\write\@auxout
	{\string\citation{#2}}\fi
\def\@citea{}\@refcite{\@for\@citeb:=#2\do
	{\@citea\def\@citea{, }\@ifundefined
	{b@\@citeb}{{\bf ?}\@warning
	{Citation `\@citeb' on page \thepage \space undefined}}
	\hbox{\csname b@\@citeb\endcsname}}}{#1}}

\def\@refcite#1#2{{#1\if@tempswa\typeout
        {IJCGA warning: optional citation argument
	ignored: `#2'} \fi}}

\def\refcite{\@ifnextchar[{\@tempswatrue
	\@refcitex}{\@tempswafalse\@refcitex[]}}

\def\pmb#1{\setbox0=\hbox{#1}
	\kern-.025em\copy0\kern-\wd0
	\kern.05em\copy0\kern-\wd0
	\kern-.025em\raise.0433em\box0}

\def\fnt#1#2{\footnotetext{\kern-.3em
	{$^{\mbox{\scriptsize #1}}$}{#2}}}

\font\tenrm=cmr10
\font\tenit=cmti10 
\font\tenbf=cmbx10
\font\bfit=cmbxti10 at 10pt
\font\ninerm=cmr9
\font\nineit=cmti9

\font\eightrm=cmr8

\def\qed{\hbox{${\vcenter{\vbox{			
   \hrule height 0.4pt\hbox{\vrule width 0.4pt height 6pt
   \kern5pt\vrule width 0.4pt}\hrule height 0.4pt}}}$}}

\begin{document}



\normalsize\textlineskip
\thispagestyle{empty}
\setcounter{page}{1}

\copyrightheading{}			

\vspace*{0.88truein}

\centerline{\bf {\Large q}-DEFORMING THE SYNCHROTRON SHAPE FUNCTION}
\vspace*{0.035truein}
\vspace*{0.37truein}
\centerline{\footnotesize \underline{HARET C. ROSU} and JOSE SOCORRO}
\vspace*{0.015truein}
\centerline{\footnotesize\it Instituto de F\'{\i}sica,
Universidad de Guanajuato, Apdo Postal E-143, Le\'on, Gto, Mexico}
\baselineskip=10pt
\vspace*{10pt}
\vspace*{0.225truein}

\vspace*{0.21truein}
\abstracts{We replace the usual integral in the shape function of the
synchrotron spectrum by a Jackson (q-deformed) integral and write down
the formulas required to calculate the Jackson first deformed form of
the synchrotron shape function.}{}{}


\textlineskip                  
\vspace*{12pt}                 

\vspace*{1pt}\textlineskip	
\vspace*{-0.5pt}
\noindent


\noindent




\noindent


Synchrotron radiation\cite{1}, first observed in 1947,
is an extremely important phenomenon in the realm of physics.
It is a nonthermal (magnetobremsstrahlung) radiation pattern, which can
be encountered in many cyclic accelerators and in
much wider astrophysical contexts \cite{td}.
On the other hand, the interesting elaborations on
the quantum deformed (basic) calculus are well established in the
mathematical literature \cite{B},
and over the past years there has been much interest
to apply the q-deformed techniques to physical phenomena and theories.
The purpose of this work is to present a q-deformation of the
synchrotron shape function.

The spectral intensity of the magnetobremsstrahlung in the synchrotron
regime is proportional to the so-called shape function \cite{1}
$$ W_{\omega}\propto
F\Big(\frac{\omega}{\omega _m}\Big)~,
\eqno(1)
$$
where $\omega _m$ is given in terms of the cyclotron radian frequency
$\omega _c$ as $\omega _m =\omega _c\; \gamma ^3$, and $F$
is given by $F(\xi) =
\frac{9\sqrt{3}}{8\pi}\xi \int _{\xi} ^{\infty}K_{5/3}(z)dz$, where $K$ is the
MacDonald function of the quoted fractional order. The small and large
asymptotic limits of the synchrotron shape function are as follows
$$
F(\xi \ll 1)\approx 1.33\;\xi ^{1/3}
\eqno(2)
$$
and
$$
F(\xi \gg 1)\approx 0.78\;\xi ^{1/2}e^{-\xi}~,
\eqno(3)
$$
with a maximum (amount of radiation)  to be found at the frequency
$\frac{1}{3}\omega _{m}$.

At the beginning of the century, F.H. Jackson \cite{j}
introduced the so called
q-integrals, which are currently known as Jackson's integrals. By definition
$$
\int _{0}^{z}f(t)d_{q}t=z(1-q)\sum _{n=0}^{\infty} f(zq^{n})q^{n}~.
\eqno(4)
$$
On the other hand, Thomae and Jackson defined a q-integral on $(0,\infty)$
by
$$
\int _{0}^{\infty}f(t)d_{q}t=(1-q)\sum_{n=-\infty}^{\infty}f(q^{n})q^{n}~.
\eqno(5)
$$
Thus, one gets
$$
\int _{z}^{\infty}f(t)d_{q}t=
\int _{0}^{\infty}f(t)d_{q}t-\int _{0}^{z}f(t)d_{q}t
\eqno(6)
$$
or
$$
\int _{z}^{\infty}f(t)d_{q}t=(1-q)\sum_{n=-\infty}^{\infty}f(q^{n})q^{n}-
z(1-q)\sum _{n=0}^{\infty} f(zq^{n})q^{n}~.
\eqno(7)
$$

In the case of the synchrotron radiation we have to take
$f$ as the q-deformed $K$ function. To get this function we can use any of the
three basic Bessel $J$ functions one can encounter in the mathematical
literature,
which are expressed in terms of the basic hypergeometric functions
$_{2}\phi _{1}$ (for the first Jackson Bessel function $J^{(J1)}$), 
$_{0}\phi _{1}$
(for the second Jackson Bessel function
$J^{(J2)}$), and $_{1}\phi _{1}$ (for the Hahn-Exton Bessel 
function $J^{(HE)}$), respectively.

Here, we shall use the first Jackson form of the q-deformed $J$, because it
does not imply the deformation of the argument as the other two basic
analogs do
(see, e.g., \cite{fv}), i.e.,
$$
J^{(J1)}_{\nu}=\frac{1}{(q;q)_{\nu}}\left(\frac{x}{2}\right)^{\nu}
\; _{2}\phi _{1}\left(0,0;q^{\nu +1};q,-\frac{x^2}{4}\right)~.
\eqno(8)
$$

From the general definition of the basic hypergeometric series
$$
_{r}\phi _{s}(a_1,a_2,..., a_{r};b_1,b_2,...,b_{s};q,x)=
\sum _{n=0}^{\infty}\frac{(a_1;q)_{n}...(a_{r};q)_{n}}{(b_1;q)_{n}...
(b_{s};q)_{n}}[(-1)^{n}q^{n(n-1)/2}]^{1+s-r}\frac{x^{n}}{(q;q)_{n}}
\eqno(9)
$$
where the q-shifted factorial symbol is defined as
$$
(a;q)_{\alpha} =\frac{(a;q)_{\infty}}{(aq^{\alpha};q)_{\infty}}~,
\eqno(10)
$$
$(a;q)_{\infty}=\prod _{k=0}^{\infty}(1-aq^{k})$, $0<q<1$, and for
$\alpha$ a positive integer $n$, $(a;q)_{n}=(1-a)(1-aq)...(1-aq^{n-1})$,
the basic hypergeometric series in the rhs of Eq.~(8) can be
calculated explicitly as follows
$$
_{2}\phi _{1}\left(0,0;q^{\nu+1};q,-\frac{x^2}{4}\right)=
\sum _{n=0}^{\infty}\left(\Big[\prod _{k=1}^{n}(1-q^{k})\Big]^{-1}
\Big[\prod _{k=0}^{n}(1-q^{\nu +1+k}\Big]^{-1}\left(\frac{-x^2}{4}\right)^{n}
\right)~.
\eqno(11)
$$

According to Ismail \cite{is} the modified q-Bessel function of the
first kind reads
$$
I^{(J1)}_{\nu}(x;q)=e^{-i\pi\nu /2}J^{(J1)}_{\nu}(ix;q)~.
\eqno(12)
$$
One can now use the well-known relation between $I_{\nu}(x)$ and 
$K_{\nu}(x)$ \cite{as}
to define basic MacDonald functions $K^{(J1)}$
$$
K^{(J1)}_{\nu}(x)=\frac{\pi}{2\sin (\nu \pi)}\Big[I^{(J1)}_{-\nu}(x)-
I^{(J1)}_{\nu}(x)\Big]~,
\eqno(13)
$$
(no deformation on the sine function!). Of course
Eq.~(13) can be applied to all three types of q-deformations.

Thus, Jackson's first q-analog of the synchrotron
shape function reads
$$
F^{(J1)}(\xi)=\frac{9\sqrt{3}}{8\pi}\xi\Big[
(1-q)\sum_{n=-\infty}^{\infty}K^{(J1)}_{5/3}(q^{n})q^{n}-
\xi (1-q)\sum _{n=0}^{\infty} K^{(J1)}_{5/3}(\xi q^{n})q^{n}\Big]
\eqno(14)
$$
and all the formulas needed to calculate $K^{(J1)}$ have been collected
herein. As $q\rightarrow 1^{-}$, $F^{(J1)}(\xi)$ goes to $F(\xi)$.

Such global types of deformation can be applied to other forms of
magnetobremsstrahlung as well, e.g., FEL ones.

\nonumsection{Acknowledgements}
\noindent
This work was partially supported by the CONACyT Projects 4868-E9406 and
3898-E9608.


\nonumsection{References}


\end{document}